\documentclass[twocolumn,aps,pre,floatfix,superscriptaddress]{revtex4-1}
\usepackage{amsmath,amssymb,eucal}
\usepackage{graphicx}
\newcommand{\pd}[2]{\frac{\partial #1}{\partial #2}}
\newcommand{\dd}[2]{\frac{d #1}{d #2}}

\begin{document}
\title{Driven Brownian coagulation of polymers}
\author{P. L. Krapivsky}
\email{pkrapivsky@gmail.com}
\affiliation {Department of Physics, Boston University, Boston, Massachusetts 02215, USA}
\author{Colm Connaughton}
\email{connaughtonc@gmail.com}
\affiliation {Mathematics Institute and Centre for Complexity Science, University of Warwick, Gibbet Hill
Road, Coventry CV4 7AL, UK}
\date{\today}


\begin{abstract}

We present an analysis of the mean-field kinetics of Brownian coagulation of droplets and polymers driven by input of monomers which aims to characterize the long time behavior of the cluster size distribution as a function of the inverse fractal dimension, $a$, of the aggregates. We find that two types of long time behavior are possible. For $0\leq a < \frac{1}{2}$ the size distribution reaches a stationary state with a power law distribution of cluster sizes having exponent $\frac{3}{2}$. The amplitude of this stationary state is determined exactly as a function of $a$. For $\frac{1}{2} < a \leq 1$,  the cluster size distribution never reaches a stationary state. Instead a bimodal distribution is formed in which a narrow population of small clusters near the monomer scale is separated by a gap (where the cluster size distribution is effectively zero) from a population of large clusters which continue to grow for all time by absorbing small clusters. The marginal case, $a=\frac{1}{2}$, is difficult to analyze definitively, but we argue that the cluster size distribution becomes stationary and there is a logarithmic correction to the algebraic tail.
\end{abstract}
\maketitle

\section{Introduction}
\label{sec-intro}

Aggregation underlies numerous phenomena from milk curdling and blood coagulation to planet and star formation. A review of the earlier work on aggregation has been given by Chandrasekhar~\cite{CHA1943}; the following work is summarized in a number of books and reviews \cite{FLO1953,DRA72,FRI2000,PK1997,ALD1999,LEY2003,KRB2010}. This research field has been initiated by a pioneering work of Smoluchowski~\cite{SMO1917} who established the framework for the analysis of aggregation.  Smoluchowski considered well-mixed diluted systems and argued that the governing equation for the concentration $c_k(t)$ of clusters of ``mass'' $k$ (that is, clusters composed of $k$ monomers, where the monomers are the clusters of the minimal mass) reads
\begin{equation}
\label{Smol}
\frac{d c_k}{dt}=\frac{1}{2}\sum_{i+j=k}K_{i,j}c_ic_j-c_k\sum_{j\geq 1}K_{k,j}c_j
\end{equation}
Here $K_{i,j}$ is the rate at which clusters of mass $i$ and $j$ merge. Only binary collisions are taken into account since the system is assumed to be diluted. Mathematically, the reaction rates $K_{i,j}$ form an infinite symmetric matrix, $K_{i,j}=K_{j,i}$; the pre-factor $\tfrac{1}{2}$ in front of the gain terms on the right-hand side of \eqref{Smol} is needed to avoid double counting. 

In addition to devising the mathematical framework underlying aggregation processes, Smoluchowski also computed the reaction rate $K_{i,j}$ in the most important case of Brownian coagulation. Namely, assuming that aggregates are spherical, Smoluchowski showed that
\begin{equation}
\label{K_Smol}
K_{i,j} = 4\pi(D_i+D_j)(R_i+R_j)
\end{equation}
where $D_k$ and $R_k$ are the diffusion coefficient and the radius of the cluster of mass $k$. Invoking the Stokes-Einstein relation between the diffusivity of a spherical (three-dimensional) object and its radius yields $D\sim R^{-1}$ and hence Eq.~\eqref{K_Smol} gives $K_{i,j}\sim (R_i^{-1}+R_j^{-1})(R_i+R_j)$.  Using additionally volume conservation, $R_k\sim k^{1/3}$, one arrives (up to a multiplicative factor that can be absorbed into the time variable) at the following expression for the Brownian reaction rate
\begin{equation}
\label{Brown}
K_{i,j}=\left(\frac{i}{j}\right)^{1/3}+\left(\frac{j}{i}\right)^{1/3}+2
\end{equation}

Rate equations \eqref{Smol} with the Brownian reaction rates \eqref{Brown} have never been solved. It is not shocking, of course, as mathematically they form an infinite set of coupled non-linear differential equations \eqref{Smol} with varying coefficients \eqref{Brown}. For simpler reaction rates, 
Eqs.~\eqref{Smol} can admit analytical solutions. For instance,  Smoluchowski already noticed~\cite{SMO1917} that the Brownian reaction rate is homogeneous, $K_{\lambda i,\lambda j}=K_{i,j}$. This feature suggests to consider the simplest such reaction rate, $K_{i,j}=\,\,$const. In this situation the rate equations \eqref{Smol} are indeed solvable~\cite{SMO1917,CHA1943}. A very few other solvable cases have been found in later work (see \cite{DRA72,ALD1999,LEY2003,KRB2010} for review), yet for all these years there has been no progress for the Brownian reaction rate. This classical case appears as analytically intractable problem as it originally looked.

In this work, we consider aggregation with input. Such driven aggregating systems often approach a non-equilibrium steady state \cite{FS1965,WHI1982,HAK1987} and the steady states tend to be more tractable. The details of input play rather limited role, e.g. they do not affect the emerging large mass behavior, if only clusters of small mass are injected. It is customary to assume that only monomers are injected. The governing equations then read 
\begin{equation}
\label{cKt}
\frac{d c_k}{dt}=\frac{1}{2}\sum_{i+j=k}K_{i,j}c_ic_j-c_k\sum_{j\geq 1}K_{k,j}c_j
+\delta_{k,1}
\end{equation}
where we set the strength of the monomer flux to unity. (The flux strength can always be absorbed into the time variable.)

In the long time limit, the mass distribution can become stationary. In this situation one must solve 
\begin{equation}
\label{cK}
0=\frac{1}{2}\sum_{i+j=k}K_{i,j}c_ic_j-c_k\sum_{j\geq 1}K_{k,j}c_j
+\delta_{k,1}
\end{equation}
The stationary solutions, when they exist, often exhibit power-law behavior. These are non-equilibrium stationary states which arise due to the flux of mass from small mass scales to large mass scales supplemented by input at the small mass scale. These states are analogous to the flux-dominated stationary states that occur in fluid turbulence \cite{FRI1995}, passive scalar advection \cite{FGV2001}, wave turbulence \cite{ZLF92}, granular gases \cite{BM2005}, and other driven aggregation systems \cite{TAK1989,KMR1998,KMR1999,CRZ2004,CRZ2005}.

In this paper we present a detailed analysis of the long-time behavior of the cluster size distribution for range of physically relevant models of Brownian coagulation with input of monomers which are parameterized by a single scaling exponent, $a$, which is related to the fractal dimension of the aggregates. We start by describing the parameter space of relevant models in Sec.~\ref{sec-models} and outline where problems of Brownian coagulation fit in this space. Next in Sec.~\ref{sec-stationary} we study the cases $0\leq a < \frac{1}{2}$ which includes Smoluchowski's original problem of Brownian coagulation of spherical droplets. We show that, in this regime, the size distribution becomes stationary for large times. For the case of spherical droplets, we show that the tail of the size distribution has the form 
\begin{equation}
\label{ck_Brown}
c_k\to C k^{-3/2}\,,\qquad C =  \frac{1}{\sqrt{4\pi}}\, \sqrt{\frac{5}{23}},
\end{equation}
in the $k\to\infty$ limit, the first exact result for this classical problem to best of our knowledge. In Sec.~\ref{sec-nonstationary} we study the cases $0\frac{1}{2} < a \leq 1$ where we find that the size distribution does not become stationary and characterize the dynamics for such kernels which include the physically interesting cases of stiff polymers ($a=1$) and polymers in an ideal solvent ($a\approx\frac{3}{5}$). Finally in Sec.~\ref{ideal}, we provide some insights into the marginal case of an ideal polymer, $a=\frac{1}{2}$ although a complete analysis eludes us. We close with a short discussion.

\section{Models of Brownian coagulation}
\label{sec-models}

Before beginning our study Brownian coagulation of polymers, let us first briefly describe the wider parameter space of models in which Brownian coagulation sits. This will help to place our subsequent work in context.

\subsection{A two-parameter class of reaction rates}
\label{ab}

The two-parameter class of homogeneous reaction rates given by
\begin{equation}
\label{Kab}
K_{i,j}=i^a j^b+i^b j^a
\end{equation}
has been extensively studied (mostly without input) in the literature. See \cite{LEY2003} for a review. 
The reason is that members of this class, for different values of the parameters $a$ and $b$, reproduce various rates (or their asymptotics) that appear in applications. 

We shall only discuss models for which the exponents $a$ and $b$ satisfy 
\begin{equation}
\label{a+b<1}
a+b\leq 1.
\end{equation}
The reason is that a system with kernel \eqref{Kab} undergoes a gelation transition when $a+b>1$.  Gelation corresponds to a loss of mass conservation in Eq.~\eqref{Smol} after a finite time \cite{ZS1980}. In the absence of a source of monomers, a gelling system reaches a steady state, but the steady state is trivial, namely the entire system is comprised of a single cluster which, in the limit of an infinite system, has infinite size. (In the presence of a source, a stationary state can be reached for such gelling systems \cite{CRZ2004} but they contain only finite total mass which considerably complicates the question of how the stationary state is approached dynamically. In particular, the scaling analysis outlined in the appendix fails since the first moment of the scaling function diverges.)

In addition, following the majority of work in this field, we shall further restrict the possible values of the parameters $a$ and $b$ to the range 
\begin{equation}
\label{ab<1}
a\leq 1, \qquad b\leq 1.
\end{equation}
The mathematical reason is that when $a>1$, and/or $b>1$, the system exhibits a highly singular phenomenon known as instantaneous gelation (see \cite{KRB2010}). This essentially means that the solution of the infinite set of rate equations is ill-defined. One can make sense of a regularized version of Eq.~\eqref{Smol} in this regime but the cut-off plays a crucial role \cite{BCSZ2011}. In any case, for systems where coagulation requires physical contact between clusters, the reaction rate cannot grow faster 
than the masses of the clusters. The requirements \eqref{ab<1} are then natural in such cases.

It is instructive to recall the derivation \cite{HAK1987} of the stationary state for the rates \eqref{Kab}, modulo the restrictions on the values of $a$ and $b$ discussed above. We shall employ the generating function approach \cite{WIL1990}. We denote generating functions by
\begin{equation}
\label{eq-GF}
\mathcal{C}_\alpha(z) = \sum_{k\geq 1} k^\alpha c_k z^k
\end{equation}
and the corresponding moments by
\begin{equation}
\label{eq-moment}
M_\alpha = \sum_{k\geq 1} k^\alpha c_k  = \mathcal{C}_\alpha(1).
\end{equation}
To find a stationary solution to \eqref{cKt}, \eqref{Kab} we multiply \eqref{cKt} by $z^k$ and sum over $k$ to obtain 
\begin{equation}
\label{AB}
\mathcal{C}_a(z)\mathcal{C}_b(z)-\mathcal{C}_a(z)M_b-\mathcal{C}_b(z)M_a+z=0.
\end{equation}
Let us assume that the cluster size distribution decays algebraically for small cluster sizes. Various exact solutions for particular kernels and the scaling argument outlined in Appendix \ref{sec-scaling} suggest that algebraic decay is plausible: 
\begin{equation}
\label{ck_asymp}
c_k\to \frac{C}{k^\tau}\qquad{\rm when}\quad k\gg 1
\end{equation}

To determine the decay exponent $\tau$ and the amplitude $C$ let us recast the conjectural asymptotic behavior of the mass distribution into a singular behavior of the generating function.  Indeed, the algebraic large $k$ behavior \eqref{ck_asymp} of $c_k$ is equivalent to the following singular behavior of the generating functions $\mathcal{C}_a(z), \mathcal{C}_b(z)$ in the $z\uparrow 1$ limit:
\begin{subequations}
\begin{align}
&\mathcal{C}_a = M_a + C\Gamma(1-\tau+a)\,(1-z)^{\tau-a-1}+\ldots
\label{Aeq}\\
&\mathcal{C}_b = M_b + C\Gamma(1-\tau+b)\,(1-z)^{\tau-b-1}+\ldots
\label{Beq}
\end{align}
\end{subequations}
Substituting expansions \eqref{Aeq}--\eqref{Beq} into equation \eqref{AB} and matching the constant terms we obtain $M_a M_b=1$. Matching then the leading $(1-z)$ behaviors we recover known results \cite{HAK1987} for the decay exponent 
\begin{equation}
\label{tau}
\tau = \frac{3+a+b}{2}\,.
\end{equation}
An expression for the amplitude is rather cumbersome, but using standard identities for the Euler's Gamma function it simplifies to   
\begin{equation}
\label{Cab}
C = \sqrt{\frac{[1-(a-b)^2]\cos[\pi(a-b)/2]}{4\pi}}
\end{equation}

\begin{figure}[tbh]
\begin{center}
\includegraphics[width=0.375\textwidth]{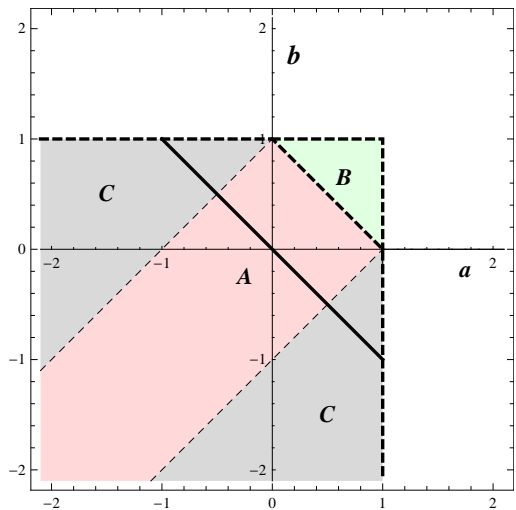}
\smallskip
\caption{\small Overview of the parameter space of the rate equations \eqref{cKt} with reaction rates \eqref{Kab}. Kernels in region A admit a steady-state solution. Kernels in region B admit a steady state solution which is preceded by a gelation transition. Kernels in region C do not admit a steady state solution. Kernels outside of these regions undergo instantaneous gelation meaning that the rate equations are ill-defined in the absence of a cut-off. The solid line $b=-a$ correspond to the family of
Brownian coagulation kernels studied in detail in this paper.}
\label{ab_fig}
  \end{center}
\end{figure}

Using Eqs.~\eqref{tau}--\eqref{Cab} we can determine when the system can reach a steady state.
First, we must assure that the series constituting $M_a$ and $M_b$ are convergent. This implies that
$\tau>1+a$ and $\tau>1+b$, which in conjunction with \eqref{tau} lead to
\begin{equation}
\label{tau1}
|a-b|<1
\end{equation}
This requirement simultaneously guarantees that the expression \eqref{Cab} for the amplitude is acceptable. The validity of \eqref{tau1} and \eqref{a+b<1} actually ensures the validity of \eqref{ab<1}.  Geometrically the region in the $(a,b)$ plane where  \eqref{tau1} and \eqref{a+b<1} are satisfied, and thus a steady state is reached, is the semi-infinite strip denoted in Fig.~\ref{ab_fig} by region A.

It is possible that the steady state is never reached. The analysis above shows that this can happen only outside the semi-infinite strip denoted by region C in Fig.~\ref{ab_fig}. For instance, a sub-monolayer epitaxial growth is described by the generalized sum-kernel reaction rates  
\begin{equation}
\label{gen_sum}
K_{i,j}=i^{-\mu}+j^{-\mu}
\end{equation}
These rates are included in the class of rates \eqref{Kab} if one sets $a=-\mu$ and $b=0$.  Two physically interesting cases are $\mu=1$ (this occurs when ad-atoms undergo ``terrace'' hopping) or $\mu=\frac{3}{2}$ (for the ``periphery'' hopping mechanism) \cite{KBE1995}. In both of these cases, and generally when 
$\mu\geq 1$, the system evolves {\em ad infinitum} \cite{KMR1998,KMR1999}, e.g. the cluster density grows as 
$(\ln t)^{\mu/2}$ for $\mu>1$. In the borderline case of $\mu=1$, there appears an additional nested logarithm \cite{KMR1998,KMR1999}, viz. the cluster density increases as $\sqrt{(\ln t)/[\ln(\ln t)]}$.  These behaviors are difficult to confirm by solving numerically mean-field rate equations, let alone to observe in simulations of a sub-monolayer epitaxial growth \cite{KA2011}.

\subsection{Brownian coagulation of polymers}
\label{polymers}

Let us now restrict our attention to the main topic of this paper, the rates for Brownian coagulation. The generalization of the Smoluchowski formula, \eqref{K_Smol}, from three to $d$ dimensions is $K_{i,j}\sim (D_i+D_j)(R_i+R_j)^{d-2}$. In the physically interesting two- and three-dimensional settings, the range of relevant models is rather limited. 

If aggregation occurs on a two-dimensional substrate \footnote{More precisely, the factor $(R_i+R_j)^{d-2}$ in Smoluchowski formula should be replaced by $1/\ln(R_i+R_j)$ in 2D; here we ignore this slowly varying logarithmic factor.}, then $K_{i,j}\sim (D_i+D_j)$. Now the diffusion coefficient often varies algebraically with the size of the cluster, $D_k\propto k^{-\mu}$ with $\mu$ being the {\em mobility} exponent, and this leads to the kernel \eqref{gen_sum}; the corresponding behaviors have been investigated in Refs.~\cite{KMR1998,*KMR1999}.

In three dimensions for spherical aggregates we arrive at \eqref{Brown}. For fractal aggregates one would get a generalized Brownian kernel 
\begin{equation}
\label{gen_Brown}
K_{i,j}=\left(\frac{i}{j}\right)^a+\left(\frac{j}{i}\right)^a+2
\end{equation}
where $a=1/D_f$ is the inverse fractal dimension of clusters. Most of the rest of this paper is devoted to the study of the solutions of Eq.~\eqref{Smol} with these rates as the parameter $a$ is varied. The only difference between Eq.~\eqref{gen_Brown} and Eq.~\eqref{Kab} with $b=-a$ is an additional constant factor in \eqref{gen_Brown} which actually {\em affects} the steady state.  The analysis, however, is similar as we shall see in the next section.  If we ignore this additional factor of 2 for the moment, the amplitude $C(a)$ given by \eqref{C} remains positive as long as $a<\frac{1}{2}$. Since $D_f=1/a$ is the fractal dimension of clusters, we conclude that the Brownian coagulation with input in three dimensions results in a stationary mass distribution if $D_f>2$. The amplitude vanishes when $D_f=2$ thereby questioning that the mass distribution reaches a stationary limit in this case. Physically $D_f=2$ is realized when clusters are membranes. 

A range of interesting behaviors arises when clusters are polymers. Coagulation often generates polymers. For instance, this naturally occurs if monomers have two reactive bonds. Schematically we can represent a monomer as $-M-$. Pairs of monomers collide to create dimers with two reactive bonds, and generally clusters are polymers with two reactive bonds: $-M=\ldots=M-$. In polymer physics, we encounter three typical situations \cite{RC2003}. For the ideal chain, equivalently a polymer in the so-called $\theta$ solvent, the polymer is essentially a random walk, so $D_f=2$, or $a=\frac{1}{2}$. For a polymer in a good solvent, $a\approx \frac{3}{5}$. Based on the discussion above, we expect that the mass distribution in this case does not reach a steady state \footnote{For the polymer in bad solvent, $a=1/3$ and we recover the classical Brownian coagulation kernel \eqref{Brown}.}. The same is true when polymers are very stiff, so that the kernel is given by  \eqref{gen_Brown} with $a=1$. 

In the following sections we address the challenge of understanding driven Brownian coagulation, as characterized by the Brownian kernel \eqref{gen_Brown}, in the range of the parameter space $0 \leq a\leq 1$ indicated by the solid line in Fig.~\ref{ab_fig}. The previous discussion suggests that the cases $a=\frac{1}{2}$, $a=\frac{3}{5}$, and $a=1$ are physically most interesting. We first study the stationary case, $0\leq a < \frac{1}{2}$ in Sec.~\ref{sec-stationary}. We then study the non-stationary case, $\frac{1}{2} < a \leq 1$, in Sec.~\ref{sec-nonstationary}. We leave the marginal case, $a=\frac{1}{2}$, to the final Sec.~\ref{ideal} since it is the most challenging.

\subsection{Remarks on numerical simulations}
\label{sec-numerics}

\begin{figure}[tbh]
\begin{center}
\includegraphics[width=0.375\textwidth]{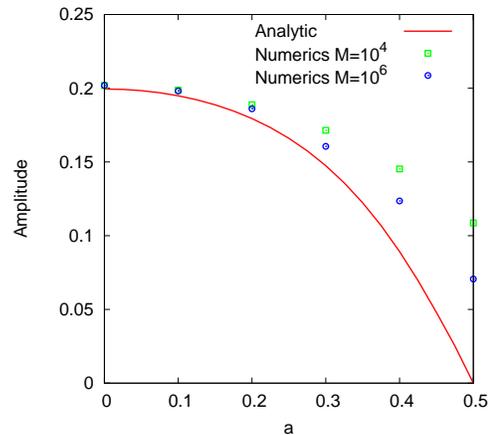}
\smallskip
\caption{Numerical measurements of the amplitude, $C$, of the stationary state 
as a function of $a$ for the kernel \eqref{Kab} with $b=-a$ using two different values for the numerical cut-off, $M$. The theoretical 
prediction, given by Eq.~(\ref{Cab}), corresponds to $M=\infty$.}
\label{fig-amplitudes}
  \end{center}
\end{figure}

All numerical simulations presented in this paper were done using a variant of the coarse-graining method described in Refs.~\cite{LEE2000,CON2009}. Numerically, one must necessarily truncate
Eq.~\eqref{Smol} at some large cluster size which we denote throughout by $M$. The choice of truncation is not unique. We chose to truncate by requiring that all clusters having mass greater than $M$ are removed from the system. For all dynamical simulations, care was taken to ensure that simulations were halted before the cluster size distribution started to feel the presence of this cut-off. We validated the code by comparing the amplitude of the stationary state against the exact result, Eq.~\eqref{Cab}. The results are shown in Fig.~\ref{fig-amplitudes} for the kernel \eqref{Kab} with $b=-a$. We can see that, our numerical measurements are in exact agreement for small values of $a$. As $a$ approaches $\frac{1}{2}$, however, there is a considerable finite size effect which diminishes as the cut-off, $M$, is increased.

\section{Stationary case: $0\leq a<\frac{1}{2}$}
\label{sec-stationary}

\subsection{Constant kernel}
\label{constant}

\begin{figure}[tbh]
\begin{center}
\includegraphics[width=0.375\textwidth]{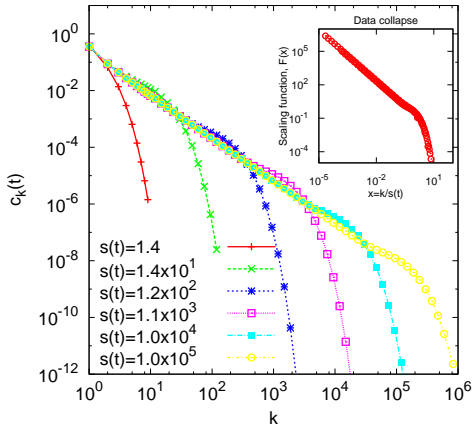}
\smallskip
\caption{Numerical calculation of the time evolution of the cluster size
distribution for the constant kernel. Inset shows the data collapse
obtained using the scaling  given by Eq.~\eqref{eq-scaling} with $\alpha=3/2$.}
\label{fig-constantKernelTimeEvolution}
  \end{center}
\end{figure}

We begin with the simplest case when the merging rate is constant, $a=0$. In this setting one can determine both the steady state mass distribution and several temporal characteristics. (This has been done e.g. \cite{FS1965,WHI1982} and reviewed in \cite{LEY2003,KRB2010}.)  Hence this model sheds the light on the emergence of the steady state.  Without loss of generality, we set $K_{i,j}=2$ in Eqs.~\eqref{ckt}. This choice can be achieved by appropriate rescaling of the density and time.
The rate equations for the aggregation process that proceeds with a constant merging rate and driven by input of monomers then read
\begin{equation}
\label{ckt}
\frac{d c_k}{dt}=\sum_{i+j=k}c_ic_j-2c_k M_0 +\delta_{k,1}.
\end{equation}
The total density of clusters, $M_0=\sum_{k\geq 1} c_k$, is just the zeroth moment of the cluster size distribution. Summing Eqs.~\eqref{ckt} we find that $M_0(t)$ satisfies 
\begin{equation}
\label{ct}
\frac{d M_0}{dt}=-M_0^2+1
\end{equation}
so for an initially empty system 
\begin{equation}
\label{ct-sol}
M_0(t)=\tanh(t)
\end{equation}
Plugging (\ref{ct-sol}) into the first rate equation  (\ref{ckt})
we obtain $\dot c_1(t)=-2\tanh(t)\,c_1(t)+1$, from which the density of
monomers is
\begin{equation}
\label{c1t-sol}
c_1(t)=\frac{1}{2}\left[\frac{t}{\cosh^2(t)}+\tanh(t)\right]
\end{equation}
Generally, one can solve Eqs.~(\ref{ckt}) recursively to give 
\begin{equation}
\label{ckt-sol}
c_k(t)=\frac{1}{\cosh^2(t)}\int_0^t dt'\,\cosh^2(t')
\sum_{i+j=k}c_i(t')c_j(t')
\end{equation}
This formal solution quickly gets very unwieldy.  Equations \eqref{ct-sol}--\eqref{c1t-sol} show that the system reaches a non-trivial steady state \footnote{We are mostly interested in the steady state, so to make formulas less cluttered we write $c_k$ instead of $c_k(\infty)$.}, $c_k(t)\to c_k$. See Fig.~\ref{fig-constantKernelTimeEvolution}.

The determination of $c_k$ from the time-dependent solution, \eqref{ckt-sol} is impractical.  A simpler procedure is based on the generating function technique introduced in Sec.~\ref{ab}.  We multiply Eq.\eqref{ckt} by $z^k$, sum over $k$ and set the left-hand side to zero to obtain a stationary state. This transforms the infinite system (\ref{ckt}) into a quadratic equation for the generating function $\mathcal{C}_0(z)$:
\begin{displaymath}
0={\cal C}_0^2-2{\cal C}_0+z.
\end{displaymath}
Solving this equation we get ${\cal C}_0(z)=1-\sqrt{1-z}$. Expanding ${\cal C}_0(z)$ we arrive at
\begin{equation}
\label{ck-sol}
c_k=\frac{1}{\sqrt{4\pi}}\,\,
\frac{\Gamma(k-\frac{1}{2})}{\Gamma(k+1)}
\end{equation}
Note that $c_k\sim k^{-3/2}$ for large $k$, so that the mass density $\sum kc_k$ diverges. This is not surprising since
\begin{equation}
\label{mass}
M_1=\sum_{k\geq 1} kc_k(t) = t
\end{equation}
due to mass conservation.  

When $t$ is large, the mass distribution $c_k(t)$ is very close to stationary for sufficiently small masses $k\ll k_*$, while for $k\gg k_*$ the mass distribution is essentially zero. The crossover mass $k_*$ is found from 
\begin{equation}
\label{mass*}
t=\sum_{k=1}^\infty kc_k(t)\approx  \sum_{k=1}^{k_*} kc_k\sim \sum_{k=1}^{k_*} k^{-1/2}\sim \sqrt{k_*}
\end{equation}
implying that $k_*\sim t^2$. The scaling analysis and numerical measurements presented in Appendix \ref{sec-scaling} suggest that $k_*\sim t^2$ is true for all values of 
$a$ in the range $0\leq a \leq 1$.

\subsection{Brownian coagulation with input}
\label{Brownian}

\begin{figure}[ht]
\begin{center}
\includegraphics[width=0.375\textwidth]{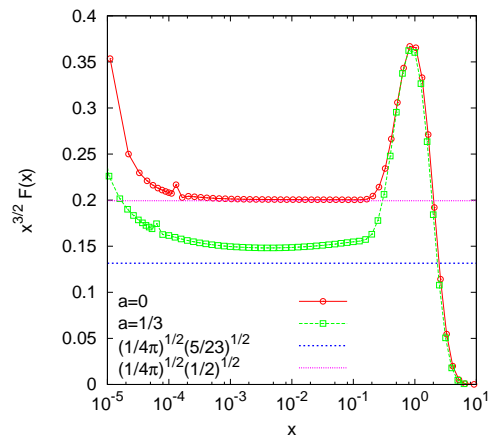}
\smallskip
\caption{Scaling functions (obtained using the scaling  given by Eq.~\eqref{eq-scaling} with $\alpha=3/2$) multiplied by $x^{3/2}$. Scaling functions for $a=0$ and $a=1/3$ are plotted. The ``bump'' phenomenon at large cluster sizes is clearly evident.}
\label{fig-bumpPhenomenon}
  \end{center}
\end{figure}
Consider steady states in a class of models with the generalized Brownian kernel \eqref{gen_Brown}. Adopting the notation of Eqs.~\eqref{eq-GF} and \eqref{eq-moment},  multiplying \eqref{cKt} by $z^k$ and summing over $k$, we arrive at the following equation for the generating functions 
\begin{eqnarray}
\nonumber 0 &=& \mathcal{C}_a(z)\mathcal{C}_{-a}(z)+\mathcal{C}_0(z)^2 - \mathcal{C}_a(z) M_{-a}-\mathcal{C}_{-a}(z) M_a\\
\label{AA} &&-2M_0\mathcal{C}_0(z)+z.
\end{eqnarray}
Expanding $\mathcal{C}_a(z)$, $\mathcal{C}_0(z)$ and  $\mathcal{C}_{-a}(z)$ in the $z\uparrow 1$ limit
\begin{equation}
\label{GF_few}
\begin{split}
&\mathcal{C}_{a}(z) = M_{a}+ C\Gamma(1-\tau-a)\,(1-z)^{\tau+a-1}+\ldots \\
&\mathcal{C}_0(z)   = M_0 +C\Gamma(1-\tau)\,(1-z)^{\tau-1}+\ldots \\
&\mathcal{C}_{-a}(z)\!=\!M_{-a}+C\Gamma(1-\tau-a)\,(1-z)^{\tau+a-1}+\ldots
\end{split}
\end{equation}
and substituting \eqref{GF_few} into Eq.~\eqref{AA} we obtain $M_a M_{-a}+M_0^2=1$ in the zeroth order 
\footnote{Thus we cannot determine the steady state cluster density $M_0$. The same holds for models considered in Sect.~\ref{ab}.}. Matching the leading terms in $(1-z)$ in this expansion, we determine the decay exponent to be $\tau=3/2$ and obtain a relation for the amplitude $C$: 
\begin{equation}
\label{C-eq}
C^2\left[\Gamma\left(a-\frac{1}{2}\right)\Gamma\left(-a-\frac{1}{2}\right)
+\Gamma^2\left(-\frac{1}{2}\right)\right]=1.
\end{equation}
From this we obtain
\begin{equation}
\label{C}
C(a) = \frac{1}{\sqrt{4\pi}}\,
\left[\frac{1}{(1-4a^2)\cos(\pi a)}+1\right]^{-1/2}.
\end{equation}
For the classical Brownian coagulation ($a=1/3$) we arrive at the announced prediction \eqref{ck_Brown} for the tail. Note that if one ignores the constant $2$ in \eqref{Brown}, then the factor $\sqrt{5/23}$ in \eqref{ck_Brown}  would be replaced by $\sqrt{5/18}$  \footnote{This follows from \eqref{Cab} with $a=1/3, b=-1/3$.}. The scaling functions (see Eq.~\eqref{eq-scaling}) for the constant kernel case, $a=0$, and the Brownian coagulation case, $a=\frac{1}{3}$, are plotted in Fig.~\ref{fig-bumpPhenomenon} multiplied by a factor of $x^\frac{3}{2}$. The plateaux for small masses indicate that the scaling exponent $\tau=\frac{3}{2}$ is very well supported numerically. Furthermore the predicted values for the amplitudes of these plateaux are in reasonable agreement.  The discrepancy for the case $a=\frac{1}{3}$ can be traced to a finite size effect. It is interesting to observe that the large mass structure of the size distribution is not trivial with a characteristic ``bump'' being clearly visible. 
We remark that a somewhat similar bump is observed hydrodynamic turbulence as 
the energy cascade enters the dissipation range. In that context, this bump
is known as the ``bottleneck effect'' \cite{FAL1994} and can be attributed to 
the depletion of nonlinear interactions due to the decay of the energy spectrum
as it enters the dissipation range. It is tempting to suggest, by analogy, that
the bump in Fig.~\ref{fig-bumpPhenomenon} is the result of the depletion of the reaction rate for large masses
due to the absence of potential coagulation partners ahead of the
front. Further investigation would be required to test this suggestion 
quantitatively. In
any case, this bump is typical for all values of $a$ and plays a very important role in the following section where we analyze the case of $a>\frac{1}{2}$.

\section{Nonstationary case $\frac{1}{2} < a \leq 1$}
\label{sec-nonstationary}

\subsection{Stiff Polymers ($a=1$)}
\label{stiff}
\begin{figure}[tb]
\begin{center}
\includegraphics[width=0.375\textwidth]{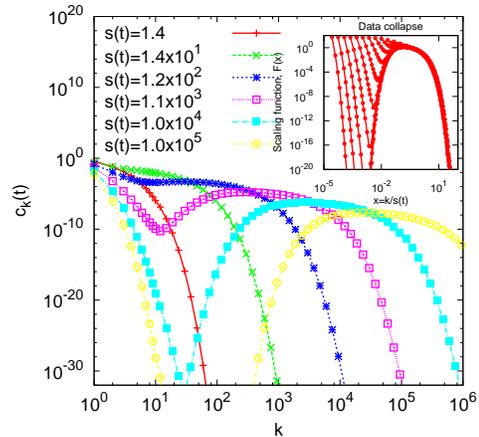}
\smallskip
\caption{Time evolution and data collapse for the case of $a=1$.}
\label{fig-stiffPolymersTimeEvolution}
  \end{center}
\end{figure}

\begin{figure}[tb]
\begin{center}
\includegraphics[width=0.375\textwidth]{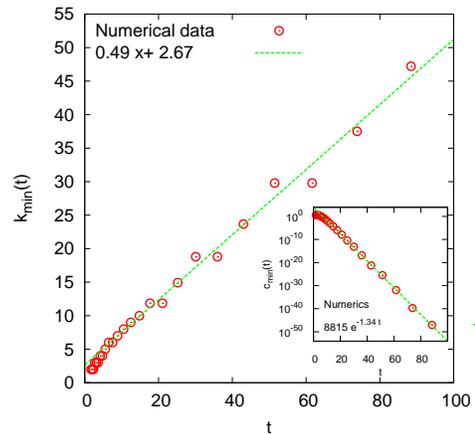}
\smallskip
\caption{Evolution of the minimum point for the case of $a=1$.}
\label{fig-stiffPolymersGap}
  \end{center}
\end{figure}

The Brownian coagulation of stiff polymers is described by reaction rates
\begin{equation}
\label{Brown_1}
K_{i,j}=\frac{i}{j}+\frac{j}{i}+2
\end{equation}

Let us first explicitly demonstrate that the stationary state is never reached. For the kernel \eqref{Brown_1}, the first few governing equations \eqref{cKt} read
\begin{equation}
\label{ckt1_few}
\begin{split}
\frac{dc_1}{dt} &= 1 - c_1(t+M_{-1}+2M_0)\\
\frac{dc_2}{dt} &= 2c_1^2 - c_2\left(\tfrac{1}{2}t+2M_{-1}+2M_0\right)\\
\frac{dc_3}{dt} &= \tfrac{9}{2}c_1 c_2 - c_3\left(\tfrac{1}{3}t+3 M_{-1}+2M_0\right)\\
\frac{dc_4}{dt} &= \tfrac{16}{3}c_1 c_3 + 2c_2^2 - c_4\left(\tfrac{1}{4}t+4M_{-1}+2M_0\right)
\end{split}
\end{equation}
where we have taken into account mass conservation \eqref{mass} and again adopted the notation \eqref{eq-moment} to denote moments of the cluster size distribution. 

In the long time limit $t=M_1\gg M_0> M_{-1}$ and therefore from \eqref{ckt1_few} we find the leading asymptotic behavior of the densities of light clusters:
\begin{equation*}
c_1=\frac{1}{t}\,,~~c_2=\frac{4}{t^3}\,,~~c_3=\frac{54}{t^5}\,,~~c_4=\frac{1280}{t^7}
\end{equation*}
These partial results reveal the general pattern, viz.
\begin{equation}
\label{ck1_asymp}
c_k = \frac{B_k}{t^{2k-1}}
\end{equation}
Thus for any fixed mass, the corresponding density approaches to zero: 
$\lim_{t\to\infty}c_k(t)=0$. The formulas \eqref{ck1_asymp} are applicable before the minimum is reached. 

Plugging \eqref{ck1_asymp} into the governing equations one finds that the amplitudes $B_k$ are determined by recurrence $B_1=1$, 
\begin{equation}
\label{Bk:rec}
B_k = \frac{k}{2}\sum_{i+j=k} K_{i,j} B_i B_j \quad\text{when}\quad k\geq 2
\end{equation}
The sequence $B_k=1, 4, 54, 1280, 44500$, etc. does not appear in  \footnote{Online Encyclopedia of Integer Sequences, http://oeis.org} and apparently does not admit a compact expression that depends only on $k$. The asymptotic behavior of the amplitudes is simple. We need only one relation, $B_k/B_{k-1}\simeq k^2$ when $k\gg 1$, which immediately follows from \eqref{Bk:rec} and \eqref{Brown_1}. Using this together with \eqref{ck1_asymp} one finds that for a fixed large $t$, the mass distribution $c_k$ quickly decreases with $k$ when $k\lesssim t$, a minimum $c_{\text{min}}\propto e^{-2t}$ is reached at $k\approx t$, and then $c_k$ starts to increase. Fig.~\ref{fig-stiffPolymersTimeEvolution} shows the time evolution of the cluster distribution for stiff polymers obtained from numerical simulations. The emergence of a minimum is clearly observed. The predicted linear increase in the position of the minimum is well supported by numerics (main panel of Fig.~\ref{fig-stiffPolymersGap}). The predicted exponential decrease of $c_{\text{min}}(t)$ with time is also well supported by numerics (inset of Fig.~\ref{fig-stiffPolymersGap}) although the rate of decrease seems to be less than 2.  

The behavior of the mass distribution in the $k\geq t$ region is harder to understand than the behavior \eqref{ck1_asymp} in the `boundary layer' region $k<t$. Before analyzing the mass distribution in the $k\geq t$ region let us look at some higher order moments which encode a lot of information about the mass distribution. Generally, the moments evolve according to rate equations 
\begin{equation}
\label{moments_few}
\begin{split}
\frac{dM_0}{dt} &= 1-\frac{1}{2}\sum_{i,j\geq 1} K_{i,j}c_ic_j\\
\frac{dM_2}{dt} &= 1+ \sum_{i,j\geq 1} K_{i,j} ij c_ic_j\\
\frac{dM_3}{dt} &= 1+ 3\sum_{i,j\geq 1} K_{i,j} i^2j c_ic_j\\
\frac{dM_4}{dt} &= 1+ 4\sum_{i,j\geq 1} K_{i,j} i^3 j c_ic_j + 3\sum_{i,j\geq 1} K_{i,j} i^2 j^2 c_ic_j
\end{split}
\end{equation}
etc. Specializing \eqref{moments_few} to the reaction rate \eqref{Brown_1} gives
\begin{equation}
\label{moments_few_1}
\begin{split}
\frac{dM_0}{dt} &= 1-(M_0^2 + t M_{-1})\\
\frac{dM_2}{dt} &= 1+ 2(M_2M_0 + t^2)\\
\frac{dM_3}{dt} &= 1+ 3(M_3M_0 + 3t M_2)\\
\frac{dM_4}{dt} &= 1+ 4M_4M_0+10M_2^2 + 14tM_3
\end{split}
\end{equation}

\begin{figure}[ht]
\begin{center}
\includegraphics[width=0.375\textwidth]{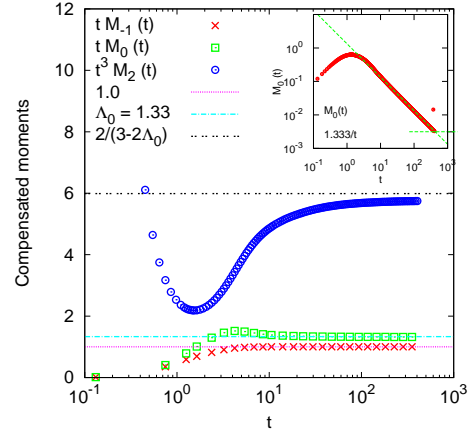}
\smallskip
\caption{Scaling of the moments with time for case $a=1$.}
\label{fig-stiffPolymersMoments}
  \end{center}
\end{figure}

We have analyzed equations \eqref{moments_few_1} using some guess work guided by simulations. The emerging behavior is quite simple. The asymptotic of the moment $M_{-1}$, and more generally any moment $M_p(t)$ with $p<0$, is dominated by monomers. In particular, the moment $M_{-1}$ which appears in Eqs.~\eqref{moments_few_1}, is given by
\begin{equation}
\label{M1t}
M_{-1} = \frac{1}{t}
\end{equation}
in the leading order (see red crosses in Fig.~\ref{fig-stiffPolymersMoments}). The moments $M_p(t)$ with $p\geq 0$ exhibit a simple scaling behavior:
\begin{equation}
\label{MLp}
M_p = \Lambda_p t^{2p-1}
\end{equation}
We have $\Lambda_1=1$ since $M_1=t$. To find other amplitudes we insert \eqref{MLp} into \eqref{moments_few_1} and obtain
\begin{equation*}
\begin{split}
3\Lambda_2    &=2(\Lambda_2 \Lambda_0 +1)\\
5\Lambda_3    &=3(\Lambda_3 \Lambda_0 + 3\Lambda_2)\\
7\Lambda_4    &=4\Lambda_4 \Lambda_0 + 10\Lambda_2^2 + 14\Lambda_3 
\end{split}
\end{equation*}
etc., from which
\begin{equation}
\label{lambdas}
\begin{split}
\Lambda_2  &=\frac{2}{\lambda}\,, \quad \lambda=3 - 2 \Lambda_0\\
\Lambda_3  &=\frac{36}{\lambda(1+3\lambda)}\\
\Lambda_4  &=\frac{40 + 624\lambda}{\lambda^2(1+3\lambda)(1+2\Lambda)}
\end{split}
\end{equation}
Generally all $\Lambda_p$ with $p\geq 2$ can be expressed through $\Lambda_0$ which therefore remains the sole unknown amplitude. Numerics (see inset of Fig.~\ref{fig-stiffPolymersMoments}) indicate that $\Lambda_0\approx \frac{4}{3}$. These relationships between moments are verified numerically for the first few moments (the main panel of Fig.~\ref{fig-stiffPolymersMoments}).

The asymptotic behaviors \eqref{M1t}--\eqref{MLp} tell us that the mass distribution exhibits a scaling behavior outside the boundary layer region. If $m$ is the typical mass scale and $c$ is the typical density scale, the moments will exhibit the scaling behavior
\begin{equation}
\label{MLp:scaling}
M_p(t) \sim c m^{p+1}
\end{equation}
The consistency between \eqref{MLp} and \eqref{MLp:scaling} sets 
\begin{equation}
\label{cm}
c\sim t^{-3}, \qquad m\sim t^2
\end{equation}
The same mass scale characterizes the Brownian coagulation in the situations when the mass distribution approaches a stationary limit, namely for the reaction rate \eqref{gen_Brown} with $|a|<1/2$. In this case, $c_k\sim k^{-3/2}$ when $k<k_*$, and the same argument as in \eqref{mass*} gives $k_*\sim t^2$.  The time evolution of the cluster size distribution is shown in Fig.~\ref{fig-stiffPolymersTimeEvolution} and clearly illustrates the predicted separation of the size distribution into two populations of clusters: a boundary layer near the monomer scale and a broader distribution of large clusters. The inset of Fig.~\ref{fig-stiffPolymersTimeEvolution} demonstrates that the large mass part of the cluster distribution collapses very well when rescaled using the theoretical scaling given by Eq.~\eqref{cm}.

In addition to the scaling $c m^{p+1}$ contribution to the moment $M_p$, there is a contribution coming from the boundary layer region where the mass distribution follows \eqref{ck1_asymp}. Actually the contribution from the boundary layer region is dominated by monomers. The scaling contribution to the moments $M_p$ dominates when $p>0$ and the monomer contribution dominates when $p<0$; at the marginal case of $p=0$ both contributions are comparable. These arguments explain \eqref{M1t}, and more generally $M_p=t^{-1}$ when $p<0$, and \eqref{MLp} for $p\geq 0$. 

Using the scaling form of the mass distribution 
\begin{equation}
\label{mass:scaling}
c_k(t) = t^{-3}F(x), \quad x=\frac{k}{t^2}
\end{equation}
we can express the amplitudes $\Lambda_p$ via the integer moments of the scales mass distribution
\begin{subequations}
\begin{align}
\label{Lambda_0}
\int_0^\infty dx\,F(x)       &= \Lambda_0 - 1\\
\label{Lambda_1}
\int_0^\infty dx\,xF(x)     &= 1\\
\label{Lambda_p}
\int_0^\infty dx\,x^pF(x) &=\Lambda_p, ~~p\geq 2
\end{align}
\end{subequations}

Let us now try to determine the scaled mass distribution $F(x)$. When $i+j=k$, the reaction rate \eqref{Brown_1} becomes $K_{i,j}=k/i+k/j$ and therefore the gain term in the Smoluchowski equation \eqref{cKt} can be re-written as
\begin{equation}
\label{gain}
\frac{1}{2}\sum_{i+j=k}K_{i,j}c_ic_j = k\sum_{j=1}^{k-1} c_{k-j}\,\frac{c_j}{j}
\end{equation}
while the loss term simplifies to 
\begin{equation}
\label{loss}
c_k\sum_{j\geq 1}K_{k,j}c_j = c_k\left(k M_{-1}+\frac{t}{k}+2M_0\right)
\end{equation}
The sum on the right-hand side of Eq.~\eqref{gain} contains the summation over the boundary layer and scaling region. The former is dominated by the contribution from monomers, while in the scaling region the summation can be replaced by integration. Therefore
\begin{equation}
\label{gain_2}
k\sum_{j=1}^{k-1} c_{k-j}\,\frac{c_j}{j} \simeq kc_1 c_{k-1} + \frac{k}{t^6} \int_0^x \frac{dy}{y} F(y) F(x-y)
\end{equation}
Inserting Eqs.~\eqref{gain}--\eqref{gain_2} into \eqref{cKt} gives
\begin{eqnarray*}
\dot c_k&=& \frac{k}{t}\,(c_{k-1}-c_k) - c_k\left(\frac{t}{k}+\frac{2\Lambda_0}{t}\right)\\
&+& \frac{k}{t^6} \int_0^x \frac{dy}{y} F(y) F(x-y)
\end{eqnarray*}
where we have also used $M_{-1}=c_1=1/t, ~M_0=\Lambda_0/t$. We now  replace the difference $c_{k-1}-c_k$ by the derivative and use the scaling form \eqref{mass:scaling} to find the governing equation for the scaled mass distribution
\begin{equation}
\label{F:eq}
F' - \frac{1-\lambda x}{x^2}\,F + \int_0^x \frac{dy}{y}\, F(y) F(x-y) = 0
\end{equation}
where, as in Eq.~\eqref{lambdas}, we used notation $\lambda=3 - 2 \Lambda_0$. We must solve \eqref{F:eq} subject to
\begin{equation}
\label{F:in}
\int_0^\infty dx\,xF(x) = 1, \quad F(0)=F(\infty)=0
\end{equation}

It seems impossible to find an analytic solution of the boundary-value problem \eqref{F:eq}--\eqref{F:in}. However, we can determine the asymptotic behaviors and the qualitative shape of the scaled mass distribution. For instance, in the small $x$ limit keeping the dominant terms simplifies \eqref{F:eq} to $F'=x^{-2}F$, from which
\begin{equation}
\label{F:small}
F \sim e^{-1/x}
\end{equation}
In the opposite $x\to\infty$ limit, the scaled mass distribution decays exponentially
\begin{equation}
\label{F:large}
F \sim x^{-\lambda} e^{-\gamma x}
\end{equation}
Plugging \eqref{F:large} into \eqref{F:eq} and taking the $x\to\infty$ limit we arrive at an equation that determines the amplitude $\gamma$:
\begin{equation*}
\gamma = \int_0^\infty \frac{dy}{y}\, F(y)\,e^{\gamma y}
\end{equation*}
Overall, the scaled mass distribution has a bell-shaped curve which vanishes exponentially fast in the $x\to 0$ and $x\to\infty$ limits. 

Note that without input the scaled mass distribution also has a bell-shaped curve and similar asymptotic behaviors \cite{VDE1985}, though the mass scale is different, $m\sim t$, and the scaling ansatz is given by $c_k(t)=t^{-2}F(k/t)$ instead of \eqref{mass:scaling}. The most important distinction from the driven case is the lack of the boundary layer region. 

\subsection{Polymers in Good Solvents: $\frac{1}{2} < a <1$}
\label{good}

\begin{figure}[ht]
\begin{center}
\includegraphics[width=0.375\textwidth]{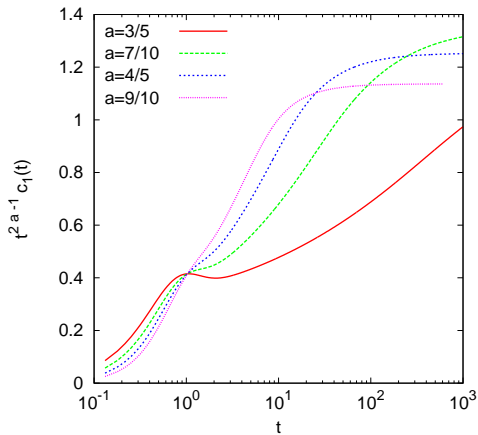}
\smallskip
\caption{Verification of scaling of the monomers density predicted by Eq.~\eqref{M0_a} for values
of $a$ in the range $\frac{1}{2} < a< 1$.}
\label{fig-goodSolventsMonomers}
  \end{center}
\end{figure}

The physically interesting situation of polymers in good solvents corresponds to $a=3/5$, but the following analysis applies to the entire region $\tfrac{1}{2}<a <1$. The procedure is essentially the same as for the $a=1$ case. The densities of monomers and dimers satisfy
\begin{equation}
\label{mon_dimer}
\begin{split}
\frac{dc_1}{dt} &= 1 - c_1(M_a+ M_{-a}+2M_0)\\
\frac{dc_2}{dt} &= 2c_1^2 - c_2\left(2^{-a} M_a+ 2^a M_{-a}+2M_0\right).
\end{split}
\end{equation}
Therefore asymptotically we have $c_1=1/M_a$ for monomers,  
$c_2=2^{1+a}c_1^2/M_a=2^{1+a}/M_a^3$ for dimers, and generally 
\begin{equation}
\label{ckA_asymp}
c_k = \frac{B_k}{M_a^{2k-1}}
\end{equation}
where the amplitudes $B_k$ are determined by recurrence $B_1=1$ and \eqref{Bk:rec}. The mass distribution is described by Eqs.~\eqref{ckA_asymp} in the boundary layer region. 

The first few integer moments satisfy
\begin{equation}
\label{moments_A}
\begin{split}
\frac{dM_0}{dt} &= 1-(M_0^2 + M_a M_{-a})\\
\frac{dM_2}{dt} &= 1+ 2(M_{1+a}M_{1-a} + t^2)\\
\frac{dM_3}{dt} &= 1+ 3(M_{2+a}M_{1-a} +M_{2-a}M_{1+a} + 2t M_2)
\end{split}
\end{equation}
From these equations we see again the validity of the scaling behavior \eqref{MLp} which implies Eqs.~\eqref{cm}--\eqref{mass:scaling}. Using \eqref{MLp}  with $p=a$ we conclude that $A\sim t^{2a-1}$ and therefore
\begin{equation}
\label{c1_a}
c_1 \sim t^{-(2a-1)}
\end{equation}
The same arguments as in Sec.~\ref{stiff} show that the monomers provide the dominant contribution to the total cluster density $M_0(t)$. Thus asymptotically
\begin{equation}
\label{M0_a}
M_0 = c_1 \sim t^{-(2a-1)}.
\end{equation}
These scaling laws are verified numerically in Fig.~\ref{fig-goodSolventsMonomers}. We remark that there is again a very strong finite size effect as $a$ approaches the
marginal value of $\frac{1}{2}$ so that for the physically interesting case of $a=\frac{3}{5}$ we were unable to integrate for times long enough to observe the final asymptotic scaling predicted by \eqref{M0_a} although the trend is clearly evident from the behavior of the larger values of $a$.

\section{Marginal case of ideal polymer chains: $a=1/2$}
\label{ideal}

\begin{figure}[tbh]
\begin{center}
\includegraphics[width=0.375\textwidth]{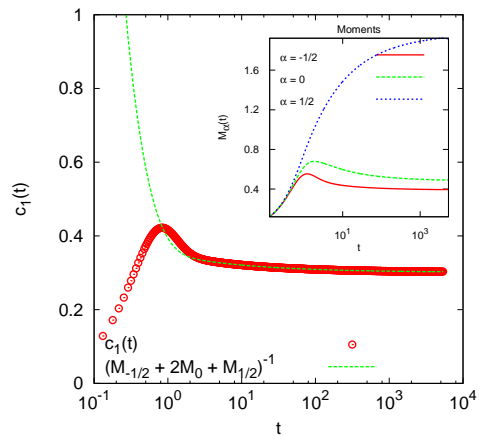}
\smallskip
\caption{Stationarity of the monomer density for $a=1/2$.}
\label{fig-idealPolymersMonomers}
  \end{center}
\end{figure}

When $a=1/2$, the amplitude $C(a)$ given by Eq.~\eqref{C} vanishes indicating that our analysis does not apply to this situation. The approach that has led to Eq.~\eqref{C} is based on two assumptions: The emergence of a stationary solution and an algebraic behavior \eqref{ck_asymp} of the stationary solution. Obviously, at least one of these assumptions is incorrect when $a=1/2$. In this marginal case the major  assumption regarding stationarity appears to work, while the secondary assumption of the power-law behavior \eqref{ck_asymp} no longer holds. Since the power-law asymptotic is valid for all $a<1/2$, one anticipates that the deviation is merely logarithmic. This suggests the conjectural asymptotic
\begin{equation}
\label{ck_log}
c_k\to C k^{-3/2}(\ln k)^{-\tau_1} \quad\text{when}\quad k\to\infty
\end{equation}
There are some parallels with a problem in wave turbulence which we studied previously \cite{CK2010}. In the turbulence setting, the corresponding expression for the
amplitude diverges rather than vanishes for a particular marginal kernel, and we did find that this divergence led to a logarithmic correction. The current problem seems to be considerably more difficult as our final asymptotic \eqref{ck_logs} involves a product over repeated logarithms. 

Before proceeding we note that the assumption that the cluster densities become stationary in the $t\to\infty$ limit tells us that the moment
\begin{equation}
\label{A_marginal}
M_\frac{1}{2} = \sum_{k\geq 1} \sqrt{k}\,c_k
\end{equation}
must remain finite. Indeed, the rate equation for the monomers 
\begin{equation*}
\frac{dc_1}{dt} = 1 - c_1(M_\frac{1}{2}+M_{-\frac{1}{2}}+2M_0)
\end{equation*}
tells us that the moments $M_\frac{1}{2}$, $M_{-\frac{1}{2}}$ and $M_0$ must remain finite (of course, the finiteness of the moment $M_\frac{1}{2}$ suffices) and the stationary value for the monomer density is given by
\begin{equation}
\label{c1_marginal}
c_1 = \frac{1}{M_\frac{1}{2}+M_{-\frac{1}{2}}+2M_0}
\end{equation}
The conjectural behavior \eqref{ck_log} of the tail is compatible with the finiteness of the moment $M_\frac{1}{2}$ when $\tau_1>1$. Since we do not know a-priori that the
cluster size distribution becomes stationary, we check the finiteness of the moments for large times numerically (see inset of Fig.~\ref{fig-idealPolymersMonomers}) and find fairly strong evidence that the relevant moments tend to finite limits as $t\to\infty$ in support of the assumption of stationarity. The stationary value of the monomer density
is well described by Eq.~\eqref{c1_marginal} as shown in the main panel of Fig.~\ref{fig-idealPolymersMonomers}.

\begin{figure}[tb]
\begin{center}
\includegraphics[width=0.37\textwidth]{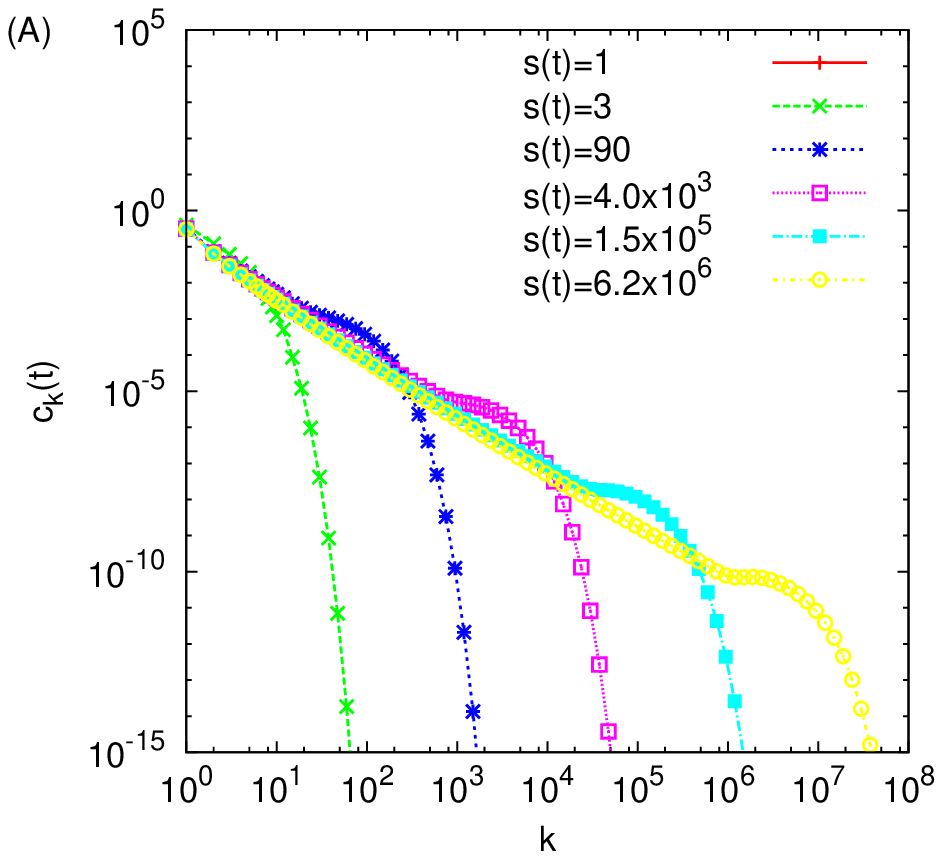}\\
\includegraphics[width=0.37\textwidth]{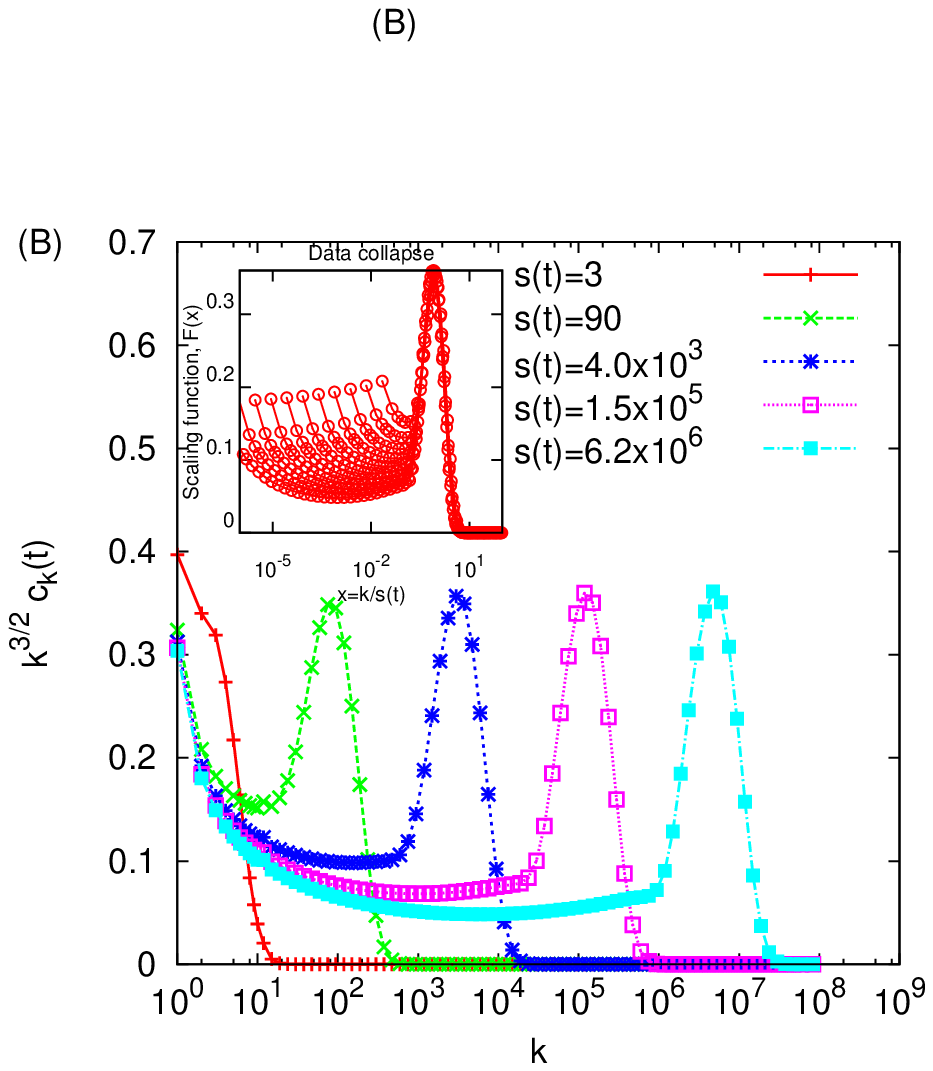}\\
\includegraphics[width=0.37\textwidth]{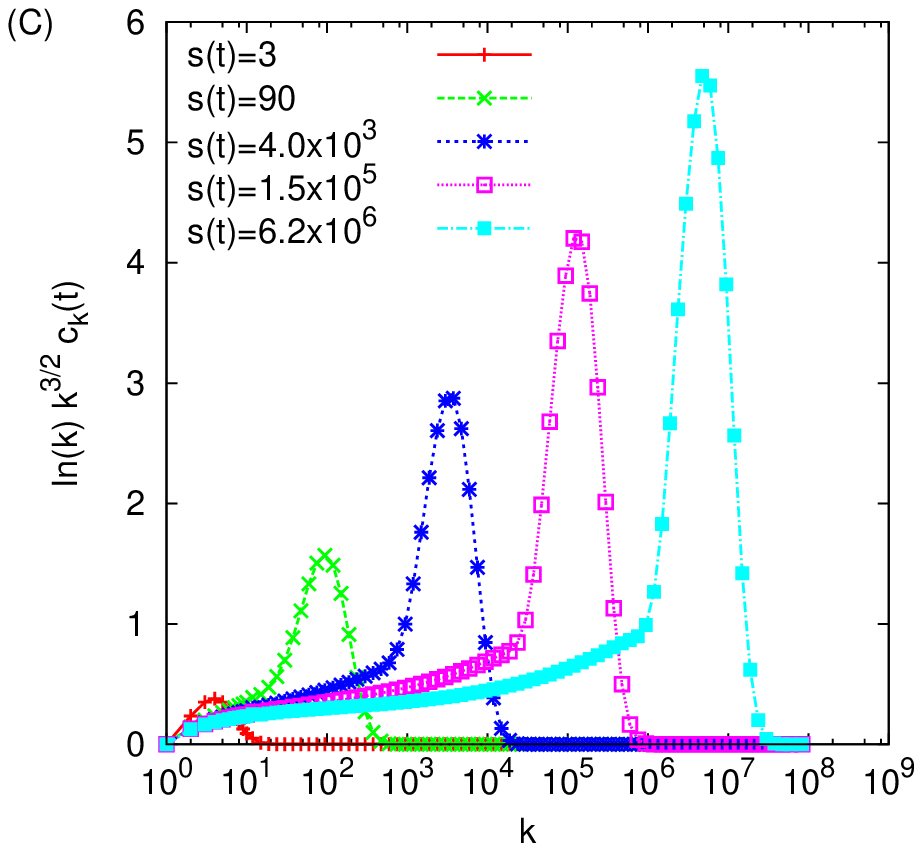}\\
\smallskip
\caption{(A) $c_k(t)$ for the case of ideal
polymers ($a=1/2$), (B) multiplied by $k^{3/2}$, (C) multiplied by $k^{3/2}\,\ln(k)$ .}
\label{fig-idealPolymers}
  \end{center}
\end{figure}

To determine the exponent $\tau_1$ we use again the generating function approach and arrive at \eqref{AA}. The conjectural behavior \eqref{ck_log} leads to certain behaviors of the generating functions in the $z\uparrow 1$ limit, yet the results are inconsistent for all $\tau_1\ne 1$. For $\tau_1=1$, the results are almost consistent, namely after canceling the dominant terms one eventually arrives at a relation of the form $1\sim \ln(\ln(1/z))$. Formally, this is of course inconsistent, although the error is much smaller than for any $\tau_1\ne 1$. Another drawback is that for $\tau_1=1$ the moment $M_\frac{1}{2}$ diverges, although in a very mild logarithmic manner. 

Thus we should take $\tau_1=1$, although the ansatz \eqref{ck_log} with this $\tau_1$ is still slightly incorrect. A similar prediction is obtained by entirely different means in Appendix \ref{sec-alternativeSS}. After a bit of trial and error one arrives at an improved ansatz 
\begin{equation}
\label{ck_log2}
c_k\to C k^{-3/2} (\ln k)^{-1} [\ln(\ln k)]^{-\tau_2}
\end{equation}
The finiteness of the moment $M_\frac{1}{2}$  now implies $\tau_2>1$. The singularity analysis of the generating function equation \eqref{AA} shows that results are inconsistent for all $\tau_2\ne 1$, while for $\tau_2=1$ the inconsistency is the weakest, namely one gets  $1\sim \ln_3(1/z)$. (We use notation $\ln_p(x)$ for the repeated logarithm: By definition $\ln_1(x)\equiv \ln(x)$ and $\ln_{p+1}(x) = \ln(\ln_p(x))$.) Thus we should take $\tau_2=1$, although the ansatz \eqref{ck_log2} with this $\tau_2$ is still slightly incorrect. Trying 
\begin{equation}
\label{ck_log3}
c_k\to C k^{-3/2} (\ln k)^{-1} (\ln_2 k)^{-1} (\ln_3 k)^{-\tau_3}
\end{equation}
one again finds that $\tau_3=1$ and the ansatz \eqref{ck_log3}  should be correct by another multiplicative factor $(\ln_4 k)^{-1}$, and then by $(\ln_5 k)^{-1}$, etc. In other words
\begin{equation}
\label{ck_logs}
c_k\to C k^{-3/2} \prod_{p\geq 1} (\ln_p k)^{-1}
\end{equation}
A similar subtle behavior characterized by an infinite product of repeated logarithms arises in reversible polymerization \cite{BK2008}. In practice, the repeated logarithms are essentially undetectable, so at best one can hope to see
\begin{equation}
\label{tail}
c_k\sim  k^{-3/2} (\ln k)^{-1} 
\end{equation}
These predictions are compared with numerics in Fig.~\ref{fig-idealPolymers}. Figure \ref{fig-idealPolymers}(A) shows the time evolution of the cluster size distribution for the marginal case $a=\frac{1}{2}$. The same data are shown in Fig.~\ref{fig-idealPolymers}(B) multiplied by a factor of $k^\frac{3}{2}$. The fact that the tail of the size distribution now looks almost flat on this plot indicates that the exponent $\tau=\frac{3}{2}$ is correct although there is still some weak variation with $k$ in the tail which may indicate a logarithmic correction. The inset of Fig.~\ref{fig-idealPolymers}(B) shows the same data collapsed according to the scaling \eqref{cm}. It is clear that, similar to what we found for the non-stationary case in Sec.~\ref{stiff}, the leading bump of the size distribution is very well described by scaling whereas the tail is not. Figure \ref{fig-idealPolymers}(C) shows the same data multiplied by an additional factor of $\log (m)$ as suggested by the theoretical arguments outlined above. We concede that it is open to interpretation whether the result demonstrates a more convincing plateau for small cluster sizes than the corresponding data without the logarithmic correction plotted in Fig.~\ref{fig-idealPolymers}(B). It seems that the honest conclusion to draw is that the numerics are unable to give definitive support to the conjectured behavior \eqref{tail} although the fact that we obtained the same answer by completely different means in Appendix \ref{sec-alternativeSS} gives us some degree of confidence that we have obtained the correct asymptotic behavior.

\section{Conclusions and discussion}
\label{sec-conclusions}

To conclude we have presented an extensive analysis of the kinetics of Brownian coagulation of polymers in the presence of a source of monomers at the mean field level. Our study focused on the determining the structure of the cluster size distribution for large times as a function of the exponent $a$, the inverse fractal dimension of the polymer aggregates. We find that this behavior falls into two classes, depending on the value of $a$.

If the exponent of the kernel is in the range $0 \leq a < \frac{1}{2}$ then the size distribution for any fixed mass becomes stationary for large times (although the typical mass continues to grow as $t^2$ for all times).  The resulting stationary state is a non-equilibrium stationary state characterized by a balance between the generation of new clusters by the injection and subsequent aggregation of monomers and the loss of clusters via aggregation. The stationary state carries a flux of mass through the space of cluster sizes from small clusters to large. We determined that the scaling exponent of this stationary state is always $\frac{3}{2}$ and obtained an exact expression for the amplitude of this stationary state as a function of $a$. During the course of this analysis, we analytically determined the stationary cluster size distribution in the presence of source of monomers for the problem of Brownian coagulation of spherical droplets ($a=\frac{1}{3}$) originally introduced by Smoluchowski at the foundation of this field almost one century ago.

If the exponent $a$ is in the range $\frac{1}{2} < a \leq 1$, which includes the physically relevant cases of stiff polymers $a=1$ and polymers in an ideal solvent $a\approx \frac{3}{5}$ we find that the cluster size distribution never reaches a stationary state. Rather the population of clusters splits into two populations resulting in a bimodal cluster size distribution. In this regime, the size distribution exhibits a narrow ``boundary layer'' of clusters near the monomer scale. This boundary layer is separated by a gap, in which the cluster size distribution goes exponentially to zero in time, from a population of large clusters which continue to grow for all times by absorbing small clusters. 

The marginal case, $a=\frac{1}{2}$ corresponding to ideal polymers, turned out to be very difficult to understand theoretically since it exhibits aspects of both behaviors. Our analysis suggests that the size distribution in this case does become stationary, but the $k^{-3/2}$ tail acquires a logarithmic correction; more work is required to make a definitive statement about the final asymptotic state for this marginal case.

\begin{acknowledgements}
CC gratefully acknowledges financial support for this research from Boston 
University and the University of Warwick. 
\end{acknowledgements}

%
\appendix

\section{Scaling analysis}
\label{sec-scaling}

\begin{figure}[t]
\begin{center}
\includegraphics[width=0.375\textwidth]{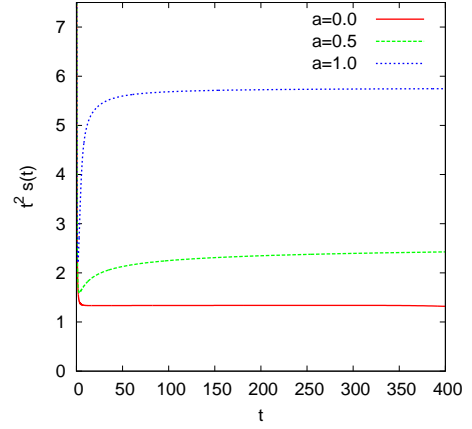}
\smallskip
\caption{Typical scale, $s(t)$, multiplied by $t^2$ for several values of $a$.}
\label{fig-typicalScale}
  \end{center}
\end{figure}

The assumption of scaling puts several strong constraints on the dynamics of the cluster size distribution which we summarize here. Consider the continuous analogue of Eq.~\eqref{Smol} obtained by replacing, the discrete cluster size index, $k$, with a continuous cluster ``mass'', $m$:
\begin{eqnarray}
\label{eq-continuousSmol} 
\dot c_m(t) &=&\frac{1}{2}
\int_0^{\infty}\int_0^{\infty}dm_1 dm_2\, K(m_1,m_2) c_{m_1}(t) c_{m_2}(t) \nonumber \\
&\times& \left[\delta(m-m_1-m_2) - \delta(m-m_1) - \delta(m-m_2)\right] \nonumber \\
&+& \delta(m-1).
\end{eqnarray}
Let us assume that the solution exhibits scaling. That is to say, there exists a monotonically increasing characteristic cluster size, $s(t)$ and a scaling function of a single variable, $F(x)$, such that the cluster size distribution tends to the the scaling form
\begin{equation}
\label{eq-scaling}
c_m(t) \to s(t)^\alpha\,F(x), \quad x=\frac{m}{s(t)}\,.
\end{equation}
Throughout this paper, we took the characteristic size to be given by the ratio of moments 
$s(t)=M_2(t)/M_1(t)$.
More precisely, the scaling behavior \eqref{eq-scaling} is supposed to emerge in the scaling limit 
$s(t)\to \infty$ and $m\to\infty$ with the scaling variable $x$ kept fixed. Let us now determine the exponent $\alpha$. Substituting Eq.~\eqref{eq-scaling} into Eq.~\eqref{eq-continuousSmol} we
obtain an integro-differential equation for the scaling function
\begin{eqnarray*}
&&\alpha F - x \dd{F}{x} = - F(x)\,\int_0^{\infty}dx_1\, K(x, x_1) F(x_1)\\
&&+\frac{1}{2}\int_0^{\infty}\int_0^{\infty}dx_1 dx_2\, K(x_1,x_2) F(x_1) F(x_2) \delta(x-x_1-x_2)
\end{eqnarray*}
and a simple differential equation for the characteristic cluster size, 
\begin{equation}
\label{eq-sdot1} 
\dot{s} = s^{\lambda +\alpha+2}\,,
\end{equation}
where $\lambda=a+b$ is the degree of homogeneity of the kernel. If we further assume that the scaling form contains the total mass:
\begin{displaymath}
\dd{}{t} \int_0^\infty m\,c_m(t)\, dm= \dd{}{t}\left[s^{\alpha+2} \int_0^\infty x\,F(x)\,dx\right]= 1
\end{displaymath}
then we obtain a second equation for $s(t)$:
\begin{equation}
\label{eq-sdot2}
\dot{s} = \frac{s^{-\alpha-1}}{F_1\,(a+2)}
\end{equation}
where $F_1 = \int_0^\infty x\,F(x)\,dx$ is the first moment of the scaling function. Comparing Eqs.~\eqref{eq-sdot1} and \eqref{eq-sdot2} we obtain the exponent $\alpha$:
\begin{equation}
\label{eq-aValue}
\alpha=\frac{\lambda+3}{2}.
\end{equation} 
Equations \eqref{eq-sdot1} and \eqref{eq-aValue} require that the typical size grows as a power law for large time:
\begin{equation}
\label{eq-s}
s(t)\sim t^\frac{2}{1-\lambda}\hspace{1.0cm}\mbox{as $t\to \infty$.}
\end{equation}
For the case of Brownian coagulation, $\lambda=0$, the assumption of scaling together with the assumption that the mass is concentrated in the scaling part of the size distribution, requires that $s(t) \sim t^2$. This prediction is in good agreement with numerics. See Fig.~\ref{fig-typicalScale} for some representative numerical results for the generalized Brownian kernel, Eq.~\eqref{gen_Brown}, for several values of $a$.

Furthermore, if the size distribution becomes stationary (and non-zero) as $t\to \infty$, then Eq.~\eqref{eq-scaling} and Eq.~\eqref{eq-aValue} require that the scaling function must be algebraic for small values of $x$ in order to cancel the time dependence:
\begin{equation}
F(x) \sim x^{-\frac{\lambda+3}{2}}\hspace{1.0cm}\mbox{as $x\to 0$.}
\end{equation}
Again for the case of Brownian coagulation, assuming scaling leads us to expect that if we have a stationary state it must scale as $k^{-3/2}$ for small cluster sizes.

\section{An alternative derivation of the stationary state}
\label{sec-alternativeSS}

An alternative way of obtaining the stationary state amplitudes given by Eq.~\eqref{Cab} or Eq.~\eqref{C} was outlined in \cite{CRZ2004}. We summarize this method here since it provides some insight into what
happens in the marginal case, $a=\frac{1}{2}$. 
If we assume that the size distribution behaves algebraically in the large $m$ limit, $c_m = C\,m^{-\tau}$,
then after some re-arrangement,  \eqref{eq-continuousSmol} can be written as
\begin{eqnarray}
\label{eq-Smol2} \dot{c}_m &=& \frac{C^2}{2}\int_0^{\infty}\int_0^{\infty}dm_1 dm_2 \\
\nonumber &&\left[  K(m_1,m_2)\, (m_1\, m_2)^{-\tau}\, \delta(m-m_1-m_2)\right.\\
\nonumber &&-K(m, m_1)\, (m\, m_1)^{-\tau}\, \delta(m_2-m-m_1)\\
\nonumber &&-  \left. K(m, m_2)\, (m\, m_2)^{-\tau}\, \delta(m_1-m_2-m)\right].
\end{eqnarray}

\begin{figure}[ht]
\begin{center}
\includegraphics[width=0.375\textwidth]{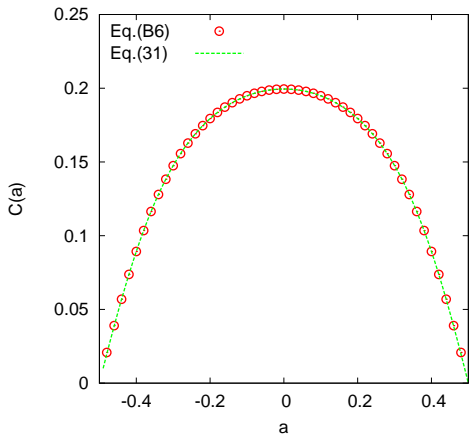}
\smallskip
\caption{Comparison of expressions \eqref{eq-Ca} and \eqref{C} for the amplitude of the stationary state.}
\label{fig-C}
  \end{center}
\end{figure}

The right-hand side contains three integrals and we now apply the following changes of variables 
\begin{eqnarray*}
\label{Z1}(m_1,m_2) &\to& \left(\frac{m m_1^\prime}{m_2^\prime}\,, ~\frac{m^2}{m_2^\prime}\right)\\
\label{Z2}(m_1,m_2) &\to& \left(\frac{m^2}{m_1^\prime}\,, ~\frac{m m_2^\prime}{m_1^\prime}\right)
\end{eqnarray*}
to the second and third integrals, respectively. After performing some 
algebra and taking advantage of the fact that the kernel is a homogeneous
function of its arguments, we obtain
\begin{equation}
\label{eq-cdot}
\dot{c}_m = \frac{1}{2}\,m^{\lambda+1-2\tau}\, C^2\,I(\tau)
\end{equation}
where
\begin{eqnarray*}
I(\tau)&=&\int_0^1 \int_0^1 d\mu_1 d\mu_2\, K(\mu_1,\mu_2)\,(\mu_1 \mu_2)^{-\tau}\\
& &\left(1-\mu_1^{2\tau-\lambda-2}-\mu_2^{2\tau-\lambda-2}\right)\,\delta(1-\mu_1-\mu_2).
\end{eqnarray*}
It is clear that a stationary state is obtained if the exponent $\tau$ is given by $\tau=(\lambda+3)/2$.
The conservation law expressing the conservation of mass,
\begin{equation}
\label{eq-massConservation}
\pd{(m\,c_m)}{t} = -\pd{J_m}{m},
\end{equation}
defines a flux of mass, $J_m$, through mass scale $m$. Multiplying 
Eq.~\eqref{eq-cdot} by $m$ and integrating therefore allows us to express
this mass flux as a function of $m$ for any value of the exponent $\tau$:
\begin{equation}
J_m(\tau) = -\frac{1}{2}\frac{C^2\,m^{\lambda+3-2\tau}\,I(\tau)}{\lambda+3-2\tau}.
\end{equation}
{}From Eq.~\eqref{eq-massConservation}, we see that the flux should become
independent of $m$ (and equal to 1 in our units) in order to have a 
stationary state.  Thus, we can evaluate the amplitude $C$ as
\begin{eqnarray}
\nonumber C &=& \lim_{\tau\to\frac{\lambda+3}{2}} \sqrt{\frac{2\,(2\tau-\lambda-3)}{m^{\lambda+3-2\tau}I(\tau)}}\\
\label{eq-C}&=& \left(\frac{1}{4}\,\left.\dd{I(\tau)}{\tau}\right|_{\tau=\frac{\lambda+3}{2}}\right)^{-\frac{1}{2}},
\end{eqnarray}
the latter step resulting from the use of  l'H\^{o}pital's rule to 
evaluate the initial undetermined expression. For the general Brownian
kernel, Eq.~\eqref{gen_Brown}, we obtain 
\begin{eqnarray}
\label{eq-Ca}
C(a) &=& \Big[-\frac{1}{2} \int_0^1\frac{d\mu}{[\mu(1-\mu)]^{3/2}}\\
\nonumber&& \left[\mu^a(1-\mu)^{-a} + (1-\mu)^a \mu^{-a} + 2\right]\\
\nonumber& &  \left[\mu\ln (\mu) + (1-\mu)\ln(1-\mu)\right]\Big]^{-\frac{1}{2}}
\end{eqnarray}
We have not succeeded in analytical computation of the integral in Eq.~\eqref{eq-Ca} and demonstration that the result is identical to \eqref{C} which was obtained via the generating function route. 
Numerical integration (Fig.~\ref{fig-C}) indicates that the two expressions are indeed 
identical. 

While this approach has yielded an answer which we already established before, it has the advantage of working for any homogeneous kernel. Furthermore, it is very helpful in providing physical insight about what happens for the marginal case, $a=\frac{1}{2}$, when $C(a)$ vanishes. If we formally repeat the preceding 
calculation taking into account the presence of a small mass cut-off at the 
monomer scale, we would obtain \footnote{This calculation is formal because we have previously assumed $C(a)$ to be independent of $m$.}
\begin{eqnarray}
\label{eq-Cam}
C(a,m) &=& \Big[-\frac{1}{2} \int_\frac{1}{m}^{1-\frac{1}{m}}\frac{d\mu}{[\mu(1-\mu)]^{3/2}}\\
\nonumber& & \left[\mu^a(1-\mu)^{-a} + (1-\mu)^a \mu^{-a} + 2\right]\\
\nonumber& &  \left[\mu\ln (\mu) + (1-\mu)\ln(1-\mu)\right]\Big]^{-\frac{1}{2}}
\end{eqnarray}
Analysis of the divergence of the integrand in Eq.~\eqref{eq-Cam} at the 
endpoints of the region of integration indicates that this divergence is
integrable if $a<\frac{1}{2}$. For such kernels, therefore, this integral
becomes independent of the monomer cut-off when we consider $m$ to be
much larger than the monomer scale and the amplitude of the stationary
state is given by a universal constant. As we have seen above, this amplitude
also characterises independence of the flux of mass through any given mass
scale, $m$, in the stationary state. This calculation tells us that when we 
consider masses much larger than the monomer mass, if $a<\frac{1}{2}$ there 
is no contribution to this flux from aggregation with monomers. The mass 
transfer is thus local in the mass space in the sense that mass is transferred
primarily through the aggregation of comparable sized clusters. This is 
contrast to nonlocal models where mass is transferred primarily through the 
aggregation of large masses with small masses of the order of the
monomer scale. From this perspective, the addition model \cite{BK1991} is
the most extreme example of non-local transfer. The corresponding analysis of 
the more general class of models given by Eq.~\eqref{Kab} shows that the mass 
flux is local in the stationary state provided $\left|a-b\right|<1$. This
is the physical origin of the region of regular stationary behaviour in 
Fig.~\ref{ab_fig}.

When $a=\frac{1}{2}$, the integrand in Eq.~\eqref{eq-Cam} diverges as $-2 \ln(\mu)/\mu$ at the lower cut-off which integrates up to give a logarithmic
dependence of $C(\frac{1}{2},m)$ on $m$ as $m$ gets large:
\begin{equation}
C(\frac{1}{2},m) \sim \frac{2}{\ln(m)}.
\end{equation}
This suggests that the stationary distribution for ideal polymers acquires a 
logarithmic correction:
\begin{displaymath}
c_k = 2\,k^{-\frac{3}{2}}\ln(k)^{-1}
\end{displaymath}
as suggested by the previous analysis of Sec. \ref{ideal}. Here it should be
noted that the assumption that $C(a,m)$ becomes independent of $m$ as
$m$ becomes large and thereby allowing it to be taken outside of the
integral in Eq.~\eqref{eq-Smol2}, remains marginally inconsistent as $m$ 
grows. Attempting to correct this inconsistency by incorporating this
logarithmic correction into Eq.~\eqref{eq-Smol2} seems likely to lead
again to the repeated logarithms of Sec. \ref{ideal}. For this reason,
although our numerics are not definitive on this matter, we believe the
evidence is in favour of stationary behaviour for the case of ideal
polymers but with logarithmic corrections to the $k^{-3/2}$ scaling of
the size distribution.
\end{document}